\documentclass{emulateapj}
\submitted{To appear in the Astrophysical Journal}

\shorttitle{SPECTROSCOPIC STUDY OF THE BHC XTE~J1118+480}
\shortauthors{TORRES ET AL.}

\begin{document}

\title{MMT Observations of the Black Hole Candidate XTE~J1118+480 near and in Quiescence\footnote{Observations reported here were obtained at the MMT Observatory, a joint facility of the University of Arizona and the Smithsonian Institution.}}
\author{M. A. P. Torres\altaffilmark{2,3},
P. J. Callanan\altaffilmark{2}, M. R. Garcia\altaffilmark{3},
P. Zhao\altaffilmark{3}, S. Laycock\altaffilmark{3}, A. K. H. Kong\altaffilmark{3}}

\altaffiltext{2}{Physics Department, University College, Cork, Ireland; mapt@phys.ucc.ie, paulc@ucc.ie}

\altaffiltext{3}{Harvard-Smithsonian Center for Astrophysics, 60 Garden St, Cambridge, MA 02138; mgarcia@cfa.harvard.edu, pzhao@cfa.harvard.edu, slaycock@cfa.harvard.edu, akong@cfa.harvard.edu}

\begin{abstract}
We report on the analysis of new and previously published MMT optical
spectra of the black hole binary XTE~J1118+480 during the decline from
the 2000 outburst to true quiescence. From cross-correlation with
template stars, we measure the radial velocity of the secondary to
derive a new spectroscopic ephemeris. The observations acquired during
approach to quiescence confirm the earlier reported modulation in the
centroid of the double-peaked H$\alpha$ emission line. Additionally,
our data combined with the results presented by Zurita et al. (2002)
provide support for a modulation with a periodicity in agreement with
the expected precession period of the accretion disk of $\sim$~52~
day. Doppler images during the decline phase of the H$\alpha$ emission
line show evidence for a hotspot and emission from the gas stream: the
hotspot is observed to vary its position, which may be due to the
precession of the disk. The data available during quiescence show that
the centroid of the H$\alpha$ emission line is offset by about
$-100$~km s$^{-1}$ from the systemic velocity which suggests that the
disk continues to precess. A H$\alpha$ tomogram reveals emission from
near the donor star after subtraction of the ring-like contribution
from the accretion disk which we attribute to chromospheric
emission. No hotspot is present suggesting that accretion from the
secondary has stopped (or decreased significantly) during
quiescence. Finally, a comparison is made with the  black hole XRN GRO
J0422+32: we show that the H$\alpha$ profile of this system also
exhibits a behaviour consistent with a precessing disk.
\end{abstract} 


\section{INTRODUCTION}
Low-Mass X-ray Binaries (LMXBs) are binary systems where mass from a
low mass secondary (M${_2}\lesssim1$~M$_\odot$) is accreted onto a
neutron star or black hole primary via an accretion disk. A subclass
of LMXBs undergo episodic outbursts, separated by periods of relative
low X-ray brightness, during which the X-ray luminosity can increase
as much as a factor $10^7$. Known as X-ray Novae (XRNe), these LMXBs
have provided us with some of the most secure dynamically confirmed stellar
black holes. For a review of the observational and theoretical status
of LMXBs, and in particular XRNe, see van Paradijs \& McClintock
(1995), Tanaka \& Shibazaki (1996), van Paradijs (1998) and McClintock
\& Remillard (2003).

XTE J1118+480 is the first identified black hole XRN located in the
Galactic halo (Wagner et al. 2001; Mirabel et al. 2001). After its
discovery during a faint (39 mCrab) X-ray outburst on March 2000
(Remillard et al. 2000), its optical counterpart was quickly
identified with a 12.9 magnitude emission-line object which had
brightened by $\sim6$ magnitudes from its pre-outburst value (Uemura
et al. 2000a; Garcia et al. 2000). The multi-wavelength outburst
spectrum of XTE J1118+480 has been intensively studied (Hynes et
al. 2000; McClintock et al. 2001a; Frontera et al. 2003). In
particular, thanks to the location of XTE J1118+480 close to the
Lockman hole it was possible to carry out observations in the extreme
UV (Hynes et al. 2000), a wavelength range inaccessible for the low
galactic latitude XRNe.

Optical photometry performed during outburst revealed a periodicity of
0.17078~d (Uemura et al. 2000b) which was later shown to be the
superhump period caused by the precession of an eccentric accretion
disk. During the decay of the outburst, McClintock et al. (2001b) and
Wagner et al. (2001) measured the radial velocity curve of the
companion star yielding to a mass function of $\sim6$~M$_\odot$. This
is beyond the maximum allowed mass for a neutron star (Rhodes \&
Ruffini 1974), a result which establishes the dynamical evidence for a
black hole primary in XTE J1118+480. They also found that the XRN is
at a distance of $1.8\pm0.6$~kpc and has a high inclination
($i\sim80^\circ$) relative to the observer. Subsequent spectroscopy
has refined the classification of the secondary to a probable
metal-poor K5-7 V subdwarf (McClintock et al. 2003). Additionally, the
outburst UV emission line spectrum showed evidence for a
nuclear-evolved donor star (Haswell et al. 2002). The mean rotational
broadening of the secondary yields a mass ratio of
$q=M{_2}/M{_1}=0.037\pm 0.007$ (Orosz 2001) which implies that the 3:2
resonant radius is within the disk tidal radius (see Whitehurst \&
King 1991 and references therein). Near quiescence, the optical light
curves displayed features characteristic of the ellipsoidal variations
of the secondary star (McClintock et al. 2001b; Wagner et al. 2001). A
12 minute offset between the spectroscopic and photometric ephemerides
was also observed and tentatively explained by Wagner et al. as due to
an asymmetrically emitting disk component. This hypothesis was
confirmed by the observations of Zurita et al. (2002) who found a
superhump component with a period 0.3 per cent longer than the orbital
period in their optical light curves obtained near quiescence. This
superhump period implies a 52 day precession period for the accretion
disk.

In a previous paper, Torres et al. (2002a) presented the results of
the analysis of their multi-epoch spectroscopy following the 2000 March
outburst. They reported variations in the double-peaked emission lines
on a time-scale longer than the orbital period of the system. This was
interpreted as evidence for the precession of an elliptical accretion
disk. During the approach of the system to quiescence, Zurita et
al. (2002) found asymmetries in the H$\alpha$ line and long-term
variations in the centroid of the line which could also be consistent with
a precessing accretion disk. Simulations of emission lines arising
from an eccentric precessing disk strongly support this interpretation
(Foulkes et al. 2004).

In this paper we re-analyze previously published data and present new
optical spectroscopy of XTE J1118+480 with the aim of studying and
resolving the complex structure of the emission line profiles. The
observations comprise data acquired at a similar time to those
presented in Zurita et al. (2002) and a data set acquired $550$
days after the system entered quiescence. The paper is structured as
follows: Section 2 presents the observations and the data reduction
procedure. In Section 3 we determine the orbital ephemeris and in
Section 4 discuss the spectrum of the donor star. Sections 5 and 6
describe the results obtained from the analysis of the Balmer emission
lines. Finally, in Section 7 we discuss our results and a summary is
given in Section 8.

\section{OBSERVATIONS AND DATA REDUCTION} 

We observed XTE J1118+480 with the Blue Channel CCD spectrograph
attached to the 6.5~m MMT at the F. L. Whipple Observatory. All the
spectra were acquired with the Loral 3072$\times$1024 CCD binned by
two in the spatial direction. Other details of the observations are
summarized in Table 1.

On six nights during the period 2000 December 1 to 2001 January 26 UT
we used a 500 grooves mm$^{-1}$ grating and a 1.25 arcsec slit width
which yielded a dispersion of 1.18~\AA~pix$^{-1}$ and a spectral
resolution of 3.6~\AA~FWHM. On night 2001 March 26 we made use instead
of a 300 grooves mm$^{-1}$ grating and a 1 arcsec slit which provided
a dispersion of 1.94~\AA~pix$^{-1}$ and a resolution of
5.5~\AA~FWHM. The source visual brightness at the beginning of the
observations was R$\sim$18.2 and it continued to fade at a rate of
0.003~mag~day$^{-1}$ \citep{zur02}. Some results obtained from the
data acquired during the first two nights can be found in McClintock
et al. (2001b). A final observing run took place on 2003 January 02/03
at a time when the system was in true quiescence (R=18.6). Spectra
were acquired using a 832 grooves mm$^{-1}$ grating and a 1.25 arcsec
slit. This combination gave a dispersion of 0.72~\AA~pix$^{-1}$ and a
resolution of 2.4~\AA~FWHM.

The images were bias and flat-field corrected with standard {\sc
iraf}\footnote{{\sc iraf} is distributed by the National Optical
Astronomy Observatories.} routines. Spectra were extracted with the
{\sc iraf kpnoslit} package. HeNeAr arc lamp exposures taken regularly
during the observations allowed us to derive the pixel-to-wavelength
calibration.  This was obtained from cubic spline fits to about 30 arc
lines giving a root-mean square deviation of typically
$\leqslant0.07$~\AA. The wavelength scales of neighbouring arc spectra
were interpolated in time and the whole process was checked and
corrected by using the atmospheric O[I] 5577.34~\AA~sky line, which
showed a rms scatter of $\leqslant9$~km s$^{-1}$ per night. Except on
2003 January 02/03, spectra of flux standards were acquired throughout
the observations and used to correct the target spectra for the
instrumental response.

\section{RADIAL VELOCITY CURVE AND ORBITAL EPHEMERIS}

We extracted the radial velocities by cross-correlating each target
spectrum with archival spectra of six template stars with spectral
types ranging through G8V-M0V and acquired using the same set-up as
for the target (see McClintock et al. 2001b). Prior to the
cross-correlation, the target and template spectra were resampled into
a common logarithmic wavelength scale and normalized by dividing with
the result of fitting a low order spline to the continuum (after
masking the emission lines in the target spectra). Cross-correlation
was carried out in the range $\lambda\lambda4900-6500$ and
least-squares sine fits with all parameters free were performed for
the radial velocity determinations. Unreliable velocities
obtained from two spectra were not included in these fits. We have
adopted the K0 V parameters of the radial velocity curve because it
provided the lowest reduced $\chi^2(=3.0)$ for the fit: $\gamma$ = 16
$\pm$ 6 km s$^{-1}$, $K_{2}$ = 709 $\pm$ 7 km s$^{-1}$,
$P=0.1699339\pm0.0000002$~d and
${T_0}=\rm{HJD}~2451880.1086\pm0.0004$, with ${T_0}$ corresponding to
the time of closest approach of the secondary to the observer. All
quoted uncertainties are 1-$\sigma$ and were obtained after increasing
the error in the radial velocities in order to give
$\chi^2_{\nu}=1$. Figure 1 displays the phase-folded radial velocity
curve. Our spectroscopic orbital period disagrees somewhat (at the
3-$\sigma$ level) with the photometric orbital period
$P{_{ph}}=0.169937\pm0.000001$~d found by Zurita et
al. (2002). Adopting the values of $K_{2}$ and the orbital period, a
refined mass function of $f(M)=6.3\pm0.2$~M$_\odot$ is
obtained. Finally, note that by using the above refined ephemeris the
outburst Doppler maps in Torres et al. (2002a) should be rotated
clockwise by $4^{\circ} \pm 1^{\circ}$, whereas for the tomograms
presented in this paper the error offsets are negligible for the first
two and within $2^{\circ}$ for the latter.

\section{THE SPECTRUM OF THE DONOR STAR}

Figure 2 presents the spectrum of XTE~J1118+480 for 2001 January
26. It was obtained by adding the spectra with different weights to
maximize the signal-to-noise ratio after Doppler-correcting them to
the rest frame of the secondary. By doing this, it is possible to see
the absorption features from the companion star that would otherwise
be blurred by the orbital motion of the star. As in McClintock et
al. (2001b) and Wagner et al. (2001), the two strongest photospheric
features detected are the NaD~{\sc i} doublet and the discontinuity in
the continuum at $\lambda$5200 caused by the Mg$b$ triplet, MgH and
(probably) the TiO$\alpha$ system. Also visible are the MgH band at
$\lambda$4780 and the molecular bands of the TiO$\gamma{^{'}}$ system
situated longwards of 5840~\AA. Metallic features such as the Ca {\sc
  i} $\lambda$4227, 6162, Fe {\sc i} $\lambda$4384 and the blend of
chiefly Fe {\sc i}+Ca {\sc i} at $\lambda$6495 are also present. The
molecular bands and lines from neutral elements identified are
characteristic of spectral-types K and M. The spectrum is bluer and
the absorption features weaker relative to the K5/8 V template shown
in the same figure. This is due to the continuum excess from the
accretion disk. In this regard, the contribution of light from the
donor star is $\sim35\pm8\%$ at $\lambda\lambda5800-6400$ on 2000
December 29 and 2001 January 26 and $60\pm10\%$ on 2003 January
02/03. These estimations were obtained by performing the standard
optimal subtraction technique (Marsh, Robinson \& Wood 1994) with
K5-M0 dwarf templates. This is comparable to values found by other
authors. In particular, the relative contribution of the secondary was
found to be $53\pm7\%$ on 2001 April (Zurita et al. 2002) and
$45\pm10\%$ on 2002 January (McClintock et al. 2003). These two
numbers along with ours from 2003 January are all consistent with a
constant $\sim55\%$ contribution from the secondary and indicate that
XTE J1118+480 has reached true quiescence on these dates. Finally,
there is no clear evidence in the night-averaged spectra for
absorption from the \hbox{Li\,{\sc i}}~$\lambda6708$ resonance line to
an equivalent width (EW) upper limit of 0.12 \AA~(1-$\sigma$)
relative to the observed continuum. In comparison, strong
\hbox{Li\,{\sc i}}~has been detected unequivocally in the companions
of other XRNe. The reader should refer to Mart\'\i n et al. (1994,
1996) for a discussion of the possible cause of the anomalously high
abundance of \hbox{Li\,{\sc i}} in XRNe.

\section{THE AVERAGED EMISSION LINE SPECTRUM}

Apart from the late-type companion spectrum, Balmer lines up to
H$\delta$ in pure emission are detected in our spectra. He{\sc i}
and Fe{\sc ii} emission lines which are commonly observed in XRNe and
DNe in quiescence are not discernible, although it has been shown that
at least He{\sc i} $\lambda5876$ is present in the spectrum of XTE
J1118+480 but veiled by the continuum from the companion star
(McClintock et al. 2003). In order to study the Balmer line profiles,
we averaged the spectra in the observer's rest-frame for the nights
when 6 or more spectra were acquired. The line profiles have a FWHM of
$\sim2500$~km s$^{-1}$~and appear double-peaked, a profile shape which
is indicative of emission arising from an accretion disk. Table 2
presents the measured full width zero intensity (FWZI) and EWs for the
lines. In Table 3 we give the position of the peaks for H$\alpha$
obtained by fitting the averaged line profiles with a 3-Gaussian
function, with one used to account for the base of the line and the
other two for the narrow peaks in the line. Also their difference
(peak-to-peak separation) and mean (centroid of the line) are
listed. From Table 2 it is obvious that the EWs measured for H$\alpha$
during 2003 January 02/03 are higher than during approach to
quiescence. This is caused in part by the lower contribution from the
accretion disk to the optical continuum during quiescence (Section
4). By looking at Table 3, there is no evidence for a change in the
peak-to-peak separation during the period of our observations, having
a weighted mean value of $1780$~km~s$^{-1}$ and a rms scatter of
60~km~s$^{-1}$. Instead, during the approach of XTE~J1118+480 to
quiescence, the position of the centroid of the line with respect to
the rest wavelength varies noticeably from night to night with an
amplitude of 250~km~s$^{-1}$. The centroid also shows displacements
during quiescence. Additionally, the averaged H$\alpha$ profiles
during the decline to quiescence show long-term changes in the
relative intensity of the red and blue peaks, a fact previously
reported by Zurita et al. (2002) and also observed during outburst
(Torres et al. 2002a). During quiescence the peaks of the line are
slightly asymmetric.

\section{THE TIME-RESOLVED H$\alpha$ LINE PROFILE}\label{tomography}

In order to study the orbital variations in the H$\alpha$ emission
line profile, we firstly determined the EWs for the individual spectra
on those nights with good orbital phase coverage. The resultant
H$\alpha$ EWs are plotted as a function of the orbital phase in Figure
1. During 2001 January 26, the EW reaches maximum at orbital phase 0.1
and no obvious minimum is observed. During 2003 January 02/03, the
H$\alpha$ EWs display a maximum at phase 0.5. The plot also shows that
the EW is significantly higher around phase 0.5 for the first night
in comparison to the second night. The variations are unlikely
to be caused by the ellipsoidal modulation in the continuum of the
secondary star since one would expect to see two equal minima in the
EWs at phases 0.25 and 0.75 corresponding to the two maxima in the
ellipsoidal modulation. The variations more probably represent
intrinsic changes in the H$\alpha$ line flux with orbital phase.

The upper panels in Figure 3 show the evolution of the H$\alpha$
profile with orbital phase in the form of trailed spectrograms. They
also illustrate the prominent long-term changes in the velocity shifts
relative to the rest wavelength of the line and in the asymmetry of
the peak intensities (Section 5). Looking to the panels, it is
possible to trace the narrow emission components present in the
double-peaked line profiles. During 2000 December 29, a sharp emission
component is visible at orbital phases 0.3-0.7 which crosses
from red to blue at phase $\sim0.4$. On 2001 January 26 the line
profile shows an emission component in the phase interval 0.7-0.9
which  crosses from blue to red at phase $\sim0.9$. During 2003
January 02/03 an emission component crosses zero velocity from red to
blue at around phase 0.5. Because of the gap around phase 0.0 in the
orbital phase coverage for the latter epoch, it is not possible to
know if the narrow component is observable at all orbital
phases.

Since the narrow components in the emission line profiles are weak and
show a complex behaviour, a multi-Gaussian fit to each individual
spectrum will render poor information about their origin. However, we
can gain insight into the H$\alpha$ emitting regions in XTE~J1118+480
by using the Doppler tomography technique (Marsh \& Horne 1988). This
technique reconstructs the brightness distribution of the binary
system in velocity space, allowing us to localize the emission
structures which are not easily recognizable in the individual
spectra. We use the maximum-entropy method (MEM) of building the
tomograms on those nights with orbital phase covering at least half an
orbital cycle and with good spectral resolution: 2000 December 29,
2001 January 26 and 2003 January 02/03. The bottom panels in Figure 3
display the derived MEM tomograms with the theoretical path of the gas
stream and the Keplerian velocities of the disk along the stream for a
radial velocity semi-amplitude of the companion star of K${_2}$=709 km
s$^{-1}$ (Section 3) and a mass ratio $q=0.037$ (Orosz 2001).

During the approach to quiescence, enhanced emission in the -V$_X$,
+V$_Y$ (upper-left) quadrant together with the emission arising from
the rotating accretion disk is easily recognizable in the Doppler
images.  The location of the former emission, halfway between the
theoretical disk and gas stream kinematic paths, is similar to that
observed in the Balmer and He{\sc ii} $\lambda4686$ maps during
outburst (Torres et al. 2002a). However, now there is a greater
overlap between this region and the theoretical trajectories. On 2000
December 29, a bright spot is present between both theoretical paths
at $0.5\pm0.1$~R$_{\mathrm{L_1}}=0.4\pm0.1~a$ where R$_{\mathrm
{L_1}}$ is the distance of the compact object from the inner
Lagrangian point and $a$ the binary separation. The position was
determined by using lines passing through points in the theoretical
paths at equal distance from ${L_1}$. On 2001 January 26 a bright spot
is again present, but now it is approximately placed on the ballistic
trajectory of the gas stream at $0.77\pm0.06$~R$_{\mathrm
{L_1}}=0.61\pm0.05~a$. Finally, note the enhanced emission in the
+V$_X$, -V$_Y$ (right-lower) quadrant for both tomograms. Its origin
will be discussed in the next section.

The tomogram computed during quiescence shows no clear evidence of
emission from a bright spot on the ring-like distribution of the
accretion disk. An emission component crossing the double-peaked
profile is evident in the trailed spectra and therefore it must be
reflected in the tomogram at a position defined by the amplitude and
phase of its modulation. In order to find its counterpart in the
tomogram, we subtracted the azimuthally symmetric part of the
map. This was obtained by taking the median of the image values within
annuli centered on the center of symmetry of the ring emission, which
falls at $\sim(-100,0)$~km~s$^{-1}$ rather than $(0,-K{_1})$. Figure 4
shows the result of the subtraction. We can see that the core of the
residual emission is situated near the predicted Roche lobe of the
secondary. The spread is approximately consistent with the combined
effect of the finite spectral resolution (110~km~s$^{-1}$) and orbital
phase smearing (ranging 20 to 270~km~s$^{-1}$).

\section{DISCUSSION}

From spectroscopic observations of XTE J1118+480 during its approach
to quiescence, Zurita et al. (2002) found that the variability in the
centroid of the nightly averaged H$\alpha$ profile was consistent with
a periodic modulation of 26 days (half of the 52 day precession period
for the accretion disk implied by the superhump period), although a 52
day periodicity was not ruled out. Combining our centroid measurements
obtained for H$\alpha$ during approach to quiescence (Table 3) with
those of Zurita et al., a modulation on the precession period is
definitely favoured and the 26 day periodicity is ruled out (Figure
5). The centroids measured for 2003 January 02/03 (Table 3) are also
displaced with respect to the systemic velocity which suggests that
the accretion disk continues precessing during true quiescence. This
is possible only if the outer disk radius is large enough to reach or
exceed the 3:2 resonance tidal instability. By using the H$\alpha$
mean peak-to-peak separation ($\Delta v$) and the semi-amplitude of
the secondary, we estimate an outer radius of $R_{\rm
d}/a=(1+q)\left(2 K_2/\Delta v\right)^2=0.66\pm0.05$ (see e.g. Dubus
et al 2001). This is a value definitely larger than the 3:2 resonance
radius r$_{32}/a(q)=0.47$ (Whitehurst \& King 1991) and comparable to
the expected disk tidal truncation radius R$_{\rm
T}$/a~=~$0.6/(1+q)=0.58$ (Warner 1995). Additionally, the hotspot in
the 2001 January 26 tomogram is located at a position consistent with
the predicted outer disk radius.

Before continuing with the discussion on the Doppler maps obtained in
the previous section, it is important to stress that the Doppler
tomography technique assumes (in particular) that all the emitting
regions co-rotate in the binary and that the flux from any of these
regions is constant with time. These assumptions might not be
fulfilled in a precessing eccentric disk during the course of a single
orbit (Foulkes et al. 2004) and therefore the results obtained from
the tomograms must be interpreted with caution. Bearing this in mind,
the tomograms for 2000 December 29 and 2001 January 26 show emission
from the hotspot and the gas stream illustrating that mass transfer
from the donor star has continued during the approach of the system to
quiescence. They also show changes in the position of the hotspot (see
Section 6). It is interesting to note here that the changes observed
in the position of the hotspot may be modulated with the precession
period because of the varying geometry of the elliptic disk. On the
other hand, the emission in the lower-right quadrant may be
interpreted as arising from an overflowing gas stream which after the
first impact intersects again the disk close to phase 0.5 (see
e.g. Kunze, Speith \& Hessman 2001) or as arising from a non-uniform
brightness distribution across the disk. In the latter case, the
formation of a region of enhanced density (temperature) in the disk
may be caused by tidal forces from the companion star (see
e.g. Steeghs \& Stehle 1999). It is clear from the long-term
variations in the centroid and peak intensities of the H$\alpha$
profile that the disk in XTE J1118+480 continues to experience the
tidal stresses from the companion star.

The residual H$\alpha$ image obtained by subtracting the symmetric
part from the 2003 January 02/03 Doppler image uncovers emission
concentrated towards the expected position of the secondary
star. H$\alpha$ emission from the hotspot or gas stream is not
detected, a fact that points to a decrease or pause in the mass
transfered from the donor star to the accretion disk. A similar
scenario was suggested to explain the quiescent H$\alpha$ tomograms of
Nova Muscae 1991 (Casares et al. 1997) and Centaurus X-4 (Torres et
al. 2002b) which show the same pattern (secondary's emission present/
absence of hotspot). In order to constrain the nature of the H$\alpha$
emission from the secondary star, we isolated the narrow line
component of the line profile as follows: first, an accretion disc
profile was created by averaging all the target spectra in the
observer's rest frame. Next, seventeen copies of the simulated disc
profile were shifted to the rest frame of the secondary star at the
times of the individual target spectra and then averaged using the
same weights as in the first step of the procedure. Finally, the
resultant profile was subtracted from the Doppler-corrected H$\alpha$
averaged profile of XTE J1118+480. This left us with a residual with
an EW=$1.0\pm0.5$~\AA~ which represents $1\%$ of the total H$\alpha$
emission. For a R quiescent magnitude of 18.6, a distance of 1.8~kpc
and assuming a secondary with a stellar radius equal to the volume
radius of its Roche lobe (R${_2}=R_{\mathrm L}(2)=0.15~a$, with
$a=2.5$~R$_\odot$ for M${_1}=7$~M$_\odot$ and $i=80^{\circ}$), we
derive from the EW of the H$\alpha$ residual a H$\alpha$ flux at the
surface of the star ({\it F}$_{{\rm H}\alpha}$) of $3\times10^{6}$ erg
cm$^{-2}$ s$^{-1}$.

Does X-ray irradiation power the H$\alpha$ emission from the
donor star? The intrinsic quiescent X-ray luminosity in the
0.3--7~keV band was estimated by McClintock et al. (2003) to be
L${_X}\approx3.5\times10^{30}$ erg s$^{-1}$. The implied X-ray
irradiation at the secondary star is then $F{_X}=L{_X}/(4\pi
a{^2})=9\times10^{6}$ erg cm$^{-2}$ s$^{-1}$ which can account for the
H$\alpha$ emission if at least a $30\%$ of the incident X-ray flux is
reprocessed to H$\alpha$ photons. We can obtain a rough estimate
of the likely fraction of the X-ray luminosity converted to H$\alpha$
photons by assuming that the H$\alpha$ emission line observed in X-ray
bright LMXBs is due completely to X-ray reprocessing in the
disk. Then:

$$ L_{\mathrm{H}\alpha}\sim
\frac{\mathrm{EW}(\mathrm{\AA})}{(7000-3000~\mathrm{\AA})}\times L_{\mathrm{opt}}
\sim 10^{-6}~L{_X}
$$

where we have taken into account that for LMXBs the averaged optical to
X-ray luminosity is L${_{\mathrm{opt}}}
(3000-7000~\mathrm{\AA})$/L${_X}(2-11~\mathrm{keV})\sim0.002$ (van
Paradijs \& McClintock 1995) and the EW of the H$\alpha$ emission line
is of the order of a few \AA~(see e.g. Shahbaz et al. 1996; Torres et
al. 2002a). Therefore we find it improbable that the X-ray irradiation could be
the source of the H$\alpha$ emission in the secondary star.

A chromospherically active secondary is an alternative explanation to
the H$\alpha$ emission. We can compare the above estimations with the
results reported by Soderblom et al. (1993) for rapidly rotating
dwarfs in the Pleiades. In order to compare the EW of the H$\alpha$
residual, we have first subtracted the underlying H$\alpha$ profile of
the secondary star from the H$\alpha$ residual by using the dwarf
templates and taking into account the veiling from the accretion disc
(Section 4). This operation yielded a corrected EW of 3~\AA. By
looking at table 6 of Soderblom et al. we find that the H$\alpha$
emission in the companion of XTE J1118+480 is comparable in strength
to that observed in some of the K dwarfs in their sample. For example,
for the K7V star listed as k403, EW=$3.41$~\AA~and {\it F}$_{{\rm
H}\alpha}=6.12\times10^{6}$ erg cm$^{-2}$ s$^{-1}$.

\subsection{Comparison of XTE J1118+480 and GRO J0422+32}

Among the black hole XRNe, GRO J0422+32 (Nova Per 1992; P=5.1~hr) has
the closest orbital period to XTE~J1118+480 and has also exhibited
superhumps during the 1992 outburst (O'Donoghue \& Charles 1996 and
references therein), suggesting a precessing accretion disk in the
system in accordance with its low mass ratio
($q=0.12^{+0.08}_{-0.07}$; Harlaftis et al. 1999). In contrast, the
decay to quiescence was interrupted by two mini-outbursts in 1993
August and December. Motivated by the similarities between these
systems, we have searched the literature for evidence of a precessing
accretion disk from the H$\alpha$ emission line of GRO J0422+32. We
have found that asymmetries in the double-peaks of the averaged line
profile have been observed (see Table 4) which we suggest are most
likely due to a precessing accretion disk. Additionally, the centroid
of the emission profiles should also vary if the disk is
precessing. This may be reflected in the H$\alpha$ systemic velocity
of $+142\pm4$~km s$^{-1}$ obtained by Garcia et al. (1996), in
comparison with the secondary's systemic velocity of
$\gamma=11\pm8$~km s$^{-1}$ (Harlaftis et al. 1999). Note that
Filippenko, Matheson \& Ho (1995) found a H$\alpha$ systemic velocity
of $26.0 \pm1.6$ ~km s$^{-1}$ by combining data obtained on 1994
November 8/9, 1995 January 26/27. Qualitatively, the H$\alpha$ and
He{\sc ii}~$\lambda4686$ trailed profiles in fig.  6 of Casares et
al. (1995) are red-shifted by $\sim100$~km s$^{-1}$. We suggest that
these variations are also associated with a precessing disk. Future
spectroscopic monitoring of XRNe in outburst and decline will enable
us to confirm unambiguously the relation between the long-term
variations in the line profiles and the precession of the accretion
disk.

\section{SUMMARY}

We have presented comprehensive optical spectroscopy of the XRN
XTE~J1118+480 from 2000 December to 2003 January. From the analysis of
the absorption line spectrum, we have established an orbital period of
$P=0.1699339\pm0.0000002$~d and a radial velocity semi-amplitude of the
secondary of $K_{2}$ = 709 $\pm$ 7 km s$^{-1}$. The implied updated
mass function is $f(M)=6.3\pm0.2$~M$_\odot$. Inspection of the
averaged spectrum of XTE J1118+480 in the rest frame of the secondary
star exhibits numerous photospheric features typical of a secondary of
spectral type K-M. Our analysis of the absorption features suggests
that the system has reached true quiescence with the companion star
contributing $\sim55\%$ of the total flux over the range
$\lambda\lambda 5800-6400$. The long-term variations in the centroid
of the H$\alpha$ profile during the decline to quiescence show a
periodicity of $\sim52$~days and an amplitude of 250 km s$^{-1}$. They
are interpreted as due to a precessing accretion disk. Comparison
between two tomograms of the H$\alpha$ line profile during the
approach of the system to quiescence show a hotspot whose position
varies. The change in the hotspot location may be related to the
changes in the gas stream impact site at the edge of the disk as it
precesses. During quiescence, both hotspot and gas stream are not
present which points to a decrease or cessation of the mass transfer
from the secondary. H$\alpha$ emission is detected near the location
of the secondary which is most probably due to chromospheric activity.


We thank Jeff McClintock who generously made available his XTE
J1118+480 data. This work was supported in part by NASA Contract
NAS8-39073 and LISA Grant NAG5-10889. Use of MOLLY, DOPPLER and
TRAILER routines developed largely by T. R. Marsh is acknowledged. We
are grateful to the anonymous referee for useful comments which
improved the quality of the manuscript.


\clearpage
\begin{figure}
\epsscale{1.0}
\plotone{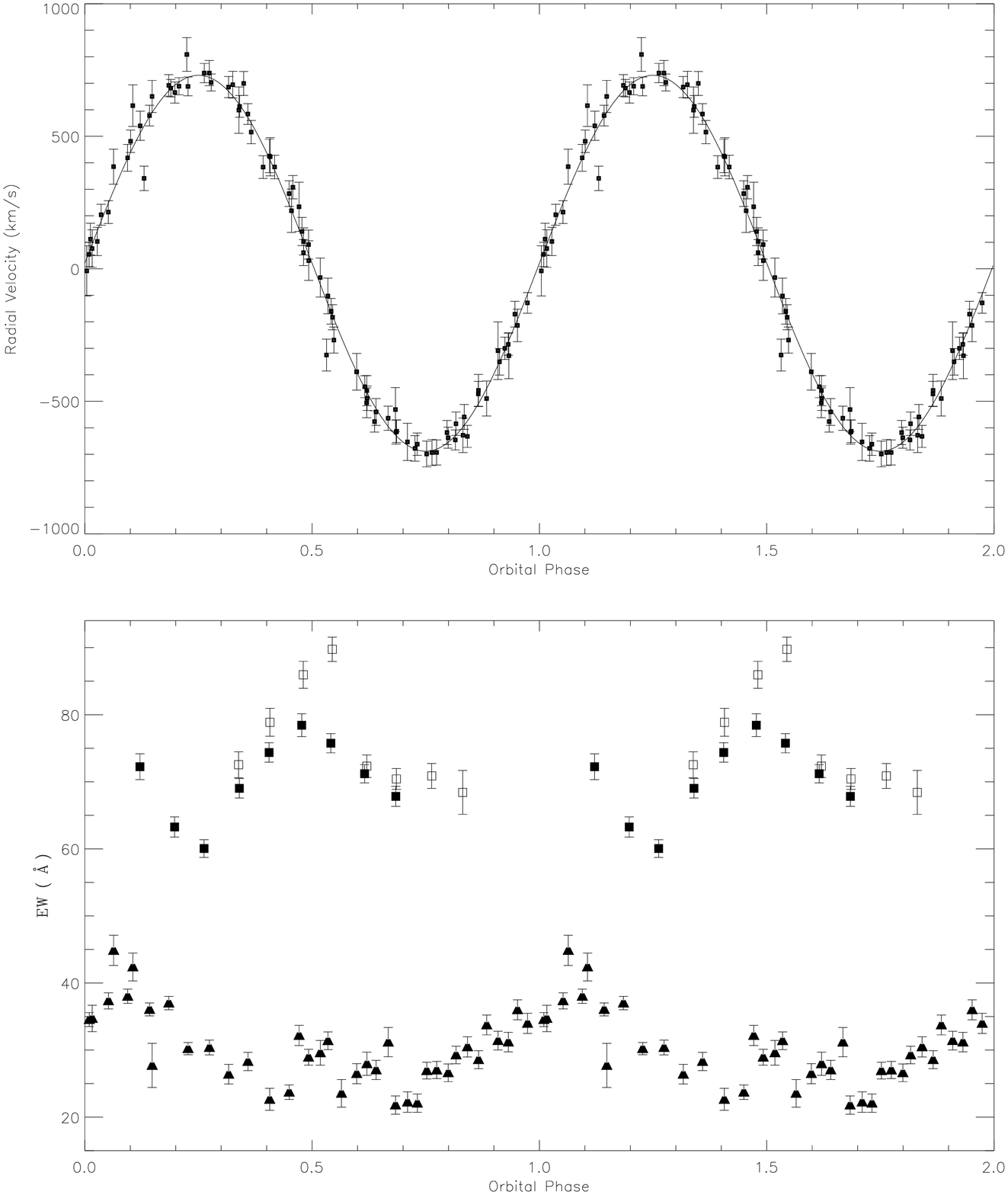}
\caption[f1.eps]{Upper panel: a total of 85 radial velocities of the secondary star
in XTE J1118+480 folded on the ephemeris of Sect. 3.1. The best sine-wave fit is also shown. Lower panel: EWs of H$\alpha$ versus orbital phase for 2001 January 26 (solid triangles), 2003 January 02 (squares) and 2003 January 03 (solid squares)}
\label{fig1}
\end{figure}

\clearpage
\begin{figure}
\epsscale{0.8}
\plotone{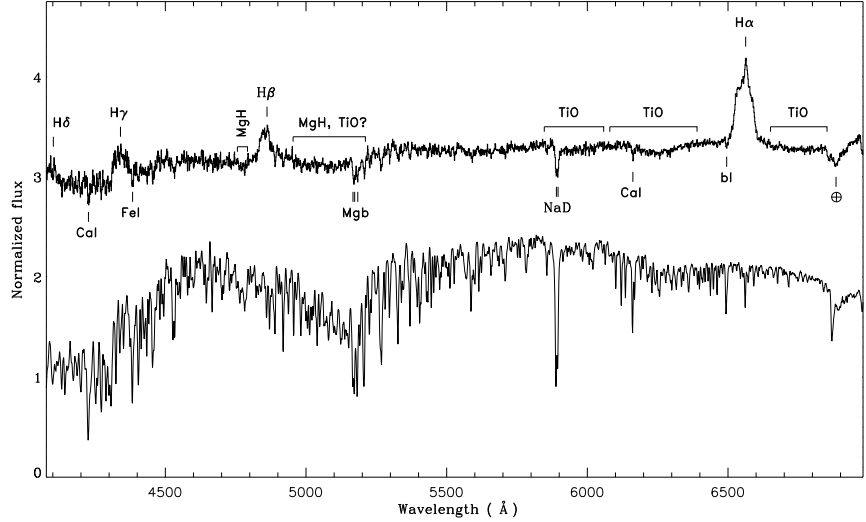}
\caption[f2.eps]{Average optical MMT spectrum of XTE J1118+480 on 2001 January 26 (UT) together with the spectrum of a K8/5 V template. The spectra have been normalized at $\lambda4100$ and the spectrum of XTE J1118+480 has been shifted vertically by +2 for the sake of clarity. Major features are identified. The Earth symbol denotes the atmospheric features and {\it bl} the metallic blend. \label{fig2}}
\end{figure}

\begin{figure}
\begin{center}
\includegraphics[angle=-90,width=5.5in]{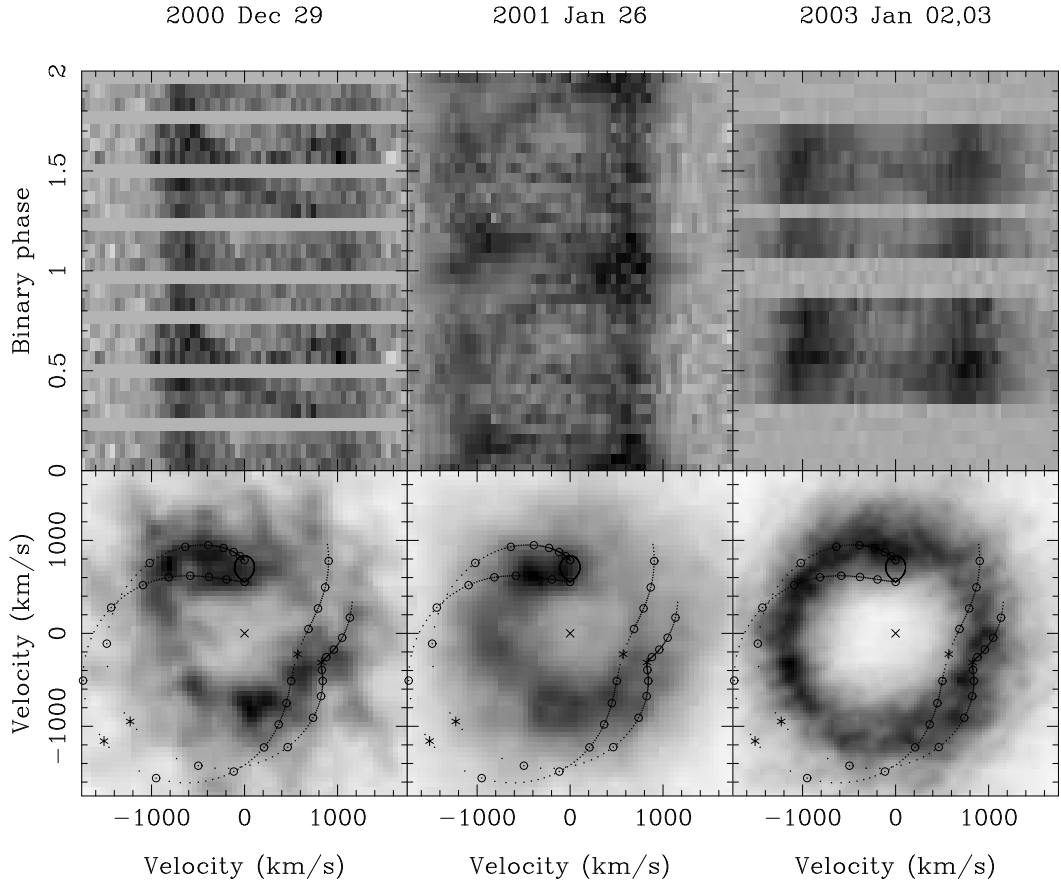}  
\caption[f3.eps]{Top panels present the trailed spectrograms of
H$\alpha$. Empty strips represent gaps in the phase coverage. For the
sake of clarity, for 2000 December 29 and 2001 January 26 the same cycle has
been plotted twice. For 2003 January, the lower and upper half of the
spectrogram are data from January 02 and 03 respectively. Lower panels
show the computed MEM Doppler maps. For 2003 January 02/03, we show the
tomogram obtained by combining the data acquired during both
nights. The Roche lobe of the secondary star, the predicted velocities
of the gas stream (lower curve) and of a Keplerian disk along the stream
(upper curve) are plotted. Distances in multiples of 0.1R$_{L1}$ are
marked along both curves with open circles. The center of mass of the
system is denoted by a cross.\label{fig3}}
\end{center}
\end{figure}

\clearpage
\begin{figure}
\begin{center}
\includegraphics[angle=-90,width=3.0in]{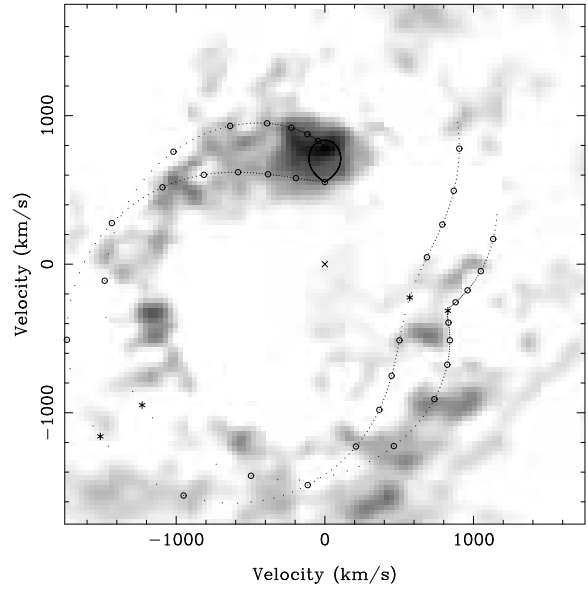}  
\caption[f4.eps]{2003 January 02/03 Doppler image of H$\alpha$ after subtracting the axisymmetric part of the emission (see text for details)\label{fig4}}
\end{center}
\end{figure}

\begin{figure}
\epsscale{1.0}
\plotone{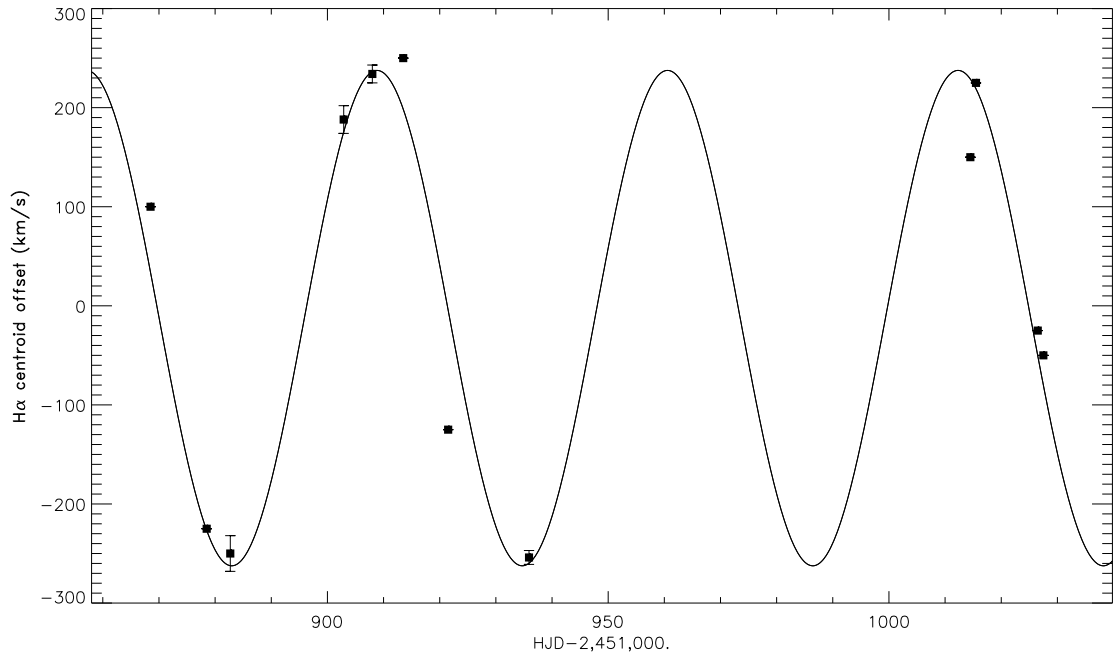}
\caption[f5.eps]{The changing centroid of the H$\alpha$ emission line during approach to quiescence. The velocity points without errors have been extracted from fig. 5 of Zurita et al. The best sine-wave fit (P$\sim52$~d) is also shown}
\label{fig5}
\end{figure}

\clearpage
\begin{table}
\begin{center}
\caption{Journal of Observations\label{Emissiontable}}
\begin{tabular}{lccccc}
\tableline\tableline
\\
{\em Date } & {\em No.} & {\em Exp. time} & 
{\em HJD start}& {\em HJD end} & {$\lambda$ range} \\
(\em UT) & spectra& {\em (s)} & (+2,451,000.)& (+2,451,000.) & (\AA) \\
\\
\tableline
\\
2000 Dec 1 	          & 2 & 900       & 880.0097 & 880.0224 & 3650-6825 \\
2000 ~~,,~ 4     		  & 6 & 900-1200  & 882.5471 & 882.9630 & 3940-7120 \\ 
2000 ~~,,~ 24   		  & 6 & 1200  & 902.8517 & 902.9851 & 4100-7015 \\
2000 ~~,,~ 28   		  & 3 & 1100-1200 & 907.0248 & 907.0543 & 4090-7005 \\
2000 ~~,,~ 29               &13 & 1200  & 907.8632 & 908.0551 & 4085-7000 \\
2001 Jan  26                &38 & 600  & 935.7534 & 936.0420 & 4075-6990 \\
2001 Mar  26   		  & 2 & 1200 & 994.7991 & 994.8147 & 3715-8510 \\
2003 Jan  02                &8 & 900 & 1,641.9798 & 1,642.0635 &4925-6850 \\
2003 Jan  03                &9 & 900 & 1,642.9625 & 1,643.0581 &4930-6860 \\

\tableline
\end{tabular}
\end{center}
\end{table}

\begin{table}
\begin{center}
\caption{FWZI$\tablenotemark{a}$~~and Equivalent Widths$\tablenotemark{a}$~~for the Balmer lines\label{fwzi}}
\begin{tabular}{lcccccccccccc}
\tableline\tableline
\\
	    & \multicolumn{2}{c}{\hrulefill~H$\alpha$ \hrulefill} & & \multicolumn{2}{c}{\hrulefill~H$\beta$ \hrulefill} & & \multicolumn{2}{c}{\hrulefill~H$\gamma$ \hrulefill} & & \multicolumn{2}{c}{\hrulefill~H$\delta$ \hrulefill} & \\
{\em UT Date } & FWZI & EW && FWZI & EW && FWZI & EW && FWZI &EW& \\
\tableline					   			      				
2000 Dec 4   & 88$\pm$5 &-22$\pm$2  && 46$\pm$10& -8$\pm$2 && 78$\pm$5 &-13$\pm$5 &&70$\pm$5 &-12$\pm$10 &\\
2000 Dec 24  & 78$\pm$5 &-27$\pm$5  && 55$\pm$6 &-12$\pm$3 && 60$\pm$5 &-9$\pm$3 &&---&---&\\
2000 Dec 29  & 85$\pm$6 &-26$\pm$4  && 65$\pm$10& -7$\pm$3 && 76$\pm$5 &-16$\pm$5 &&---&---&\\
2001 Jan 26  & 85$\pm$5 &-32$\pm$6  && 51$\pm$4 &-11$\pm$4 && 76$\pm$5 &-16$\pm$9 &&---&---&\\
2003 Jan 02  & 90$\pm$2 &-76$\pm$8  &&---       & ---      &&---       &---       &&---&---&\\
2003 Jan 03  & 99$\pm$2 &-70$\pm$6  &&---       & ---      &&---       &---       &&---&---&\\

\tableline 
\end{tabular}
\tablenotetext{a}{In angstroms.}
\tablecomments{The EWs are the mean of the values obtained for the individual profiles and the uncertainties correspond to the standard deviation. The uncertainties in the FWZIs were estimated by looking at the scatter in the values when selecting different wavelength intervals to set the local continuum level.}	
\end{center}
\end{table}    

\begin{table}
\begin{center}
\caption{Fits to the H$\alpha$ Line Profile\label{hahb}}   
\begin{tabular}{lcccccc}
\tableline\tableline
\\
{\em Date } & V$_{b}$       & V$_{r}$       & V$_{r}-$V$_{b}$ & (V$_{b}$+V$_{r}$)/2 &\\
{\em UT}    & (km s$^{-1}$) & (km s$^{-1}$) & (km s$^{-1}$)   & (km s$^{-1}$)       &\\
\tableline					     						
2000 Dec 4   &-1163$\pm$34 & 664$\pm$14  & 1827$\pm$37 &-250 $\pm$18 &\\
2000 Dec 24  & -644$\pm$16 & 1021$\pm$23 & 1665$\pm$28 & 188 $\pm$14 &\\
2000 Dec 29  &-651 $\pm$8  & 1120$\pm$17 & 1771$\pm$19 & 234 $\pm$9  &\\
2001 Jan 26  &-1134$\pm$12 &  627$\pm$7  & 1761$\pm$14 &-254 $\pm$7  &\\
2003 Jan 02  & -952$\pm$6  &  811$\pm$8  & 1763$\pm$10 &-71  $\pm$5  &\\
2003 Jan 03  &-1009$\pm$6  &  823$\pm$8  & 1832$\pm$10 &-93 $\pm$5   &\\
\tableline
\end{tabular}
\tablecomments{V$_{b}$ and V$_{r}$ designate the positions (shifts) respect to the rest wavelength of the line of the blue and red peaks respectively. The uncertainties were calculated after scaling the error bars to give ${\chi^2_\nu}=1$.}
\end{center}
\end{table}

\clearpage
\begin{table}
\begin{center}
\caption{Asymmetries in the averaged double-peaked H$\alpha$ profile of GRO J0422+32\label{Emissiontable}}
\begin{tabular}{lcclc}
\tableline\tableline
\\
Night/s      & Net Exposure & Stronger  & Comments              & refs. \\  
averaged     & (hr)         & peak      &                       &        \\
\tableline
\\
1993 Oct 10    & 5.3          &     Blue      & Decay to quiescence after August mini-outburst              & 1 \\	
1993 Oct 12    & 5.3          &     Blue      &                                  &   \\
1993 Dec 9     & 6     	      &     Red       & Maximum December mini-outburst   & 2 \\	
1993 Dec 10    & 5     	      &     Red       &                                  &   \\ 
1993 Dec 16-19 & $\sim11$     &	    Blue      & Initial decline of mini-outburst & 3 \\
1994 Nov 5-7   & $>14.5$      &     Blue      & Quiescence                       & 4 \\
1995 Jan 26    & 4.9          &     Red       & Quiescence                       & 5 \\

 \tableline
 \end{tabular}
 \tablecomments{References: 1.- Garcia et al. (1996); 2.- fig. 5d in Callanan et al. (1995); 3.- fig. 2 of Casares et al. (1995); 4.- fig. 2 in Orosz \& Bailyn (1995); 5.- fig. 1 in Harlaftis et al. (1999)} 
\end{center}
\end{table}


\clearpage
\begin{thebibliography}{}


\bibitem[Callanan et al.(1995)]{call95}Callanan, P. J., et al. 1995, ApJ, 441, 786

\bibitem[Casares et al.(1995)]{cas95}Casares, J, Charles, P. A., \& Marsh, T. R. 1995, MNRAS, 277, 45

\bibitem[Casares et al.(1997)]{cas97}Casares, J., Mart\'\i{}n, E. L.,
Charles, P. A., Molaro, P., \& Rebolo, R. 1997, NewA, 1, 299

\bibitem[Dubus et al.(2001)]{dub01}Dubus, G., Kim Rita, S. J., Menou, K., Szkody, P., \& Bowen, D. V. 2001, ApJ, 553, 307

\bibitem[Filippenko et al.(1995)]{fil95}Filippenko, A. V., Matheson, T., \& Ho, L. C. 1995, \apj, 455, 614

\bibitem[Foulkes et al.(2004)]{fou04}Foulkes, S. B., Haswell, C. A., Murray, J. R., \& Rolfe, D. 2004, \mnras, 349, 1179

\bibitem[Frontera et al.(2003)]{fro03}Frontera, F., et al. 2003, ApJ, 592, 1110

\bibitem[Garcia et al.(1996)]{gar96}Garcia, M. R., Callanan, P. J., McClintock, J. E., \& Zhao, P. 1996, \apj, 460, 932

\bibitem[Garcia et al.(2000)]{gar00}Garcia, M., Brown, W. Pahre, M., McClintock, J., Callanan, P., \& Garnavich, P. 2000, \iaucirc~7392

\bibitem[Harlaftis et al.(1999)]{har99}Harlaftis, E. T., Collier, S.,
Horne, K., \& Filippenko, A. V. 1999, A\&Ap, 341, 491

\bibitem[Haswell et al.(2002)]{has02}Haswell, C. A., Hynes, R. I., King, A. R., \& Schenker, K. 2002, \mnras, 332, 928

\bibitem[Hynes et al.(2000)]{hyn00}Hynes, R. I., Mauche, C. W., Haswell, C. A., Shrader, C. R., Cui, W., \& Chaty, S. 2000, \apj, 539, 37

\bibitem[Kunze et al.(2001)]{kun01}Kunze, S., Speith, R, \& Hessman, F. V. 2001, \mnras, 322, 499

\bibitem[Marsh \& Horne(1988)]{mar88}Marsh, T.R., \& Horne, K. 1988, \mnras, 235, 269  

\bibitem[Marsh, Robinson \& Wood(1994)]{mar94}Marsh, T.R., Robinson, E. L., \& Wood, J. H. 1994, \mnras, 266, 137

\bibitem[Mart\'\i n et al.(1994)]{mart94}Mart\'\i n, E. L., Rebolo,
R., Casares, J., \& Charles, P. A. 1994, ApJ, 435, 791

\bibitem[Mart\'\i n et al.(1996)]{mart96}Mart\'\i n, E. L., Casares,
J., Molaro, P., Rebolo, R., \& Charles P. 1996, NewA, 1, 197

\bibitem[McClintock et al.(2001a)]{mcc01B}McClintock, J. E., et al. 2001a, \apj, 555, 477

\bibitem[McClintock et al.(2001)]{mcc01A}McClintock, J. E., Garcia,  M. R., Caldwell, N., Falco, E. E., Garnavich, P. M., \& Zhao, P. 2001b, \apjl, 551, 147L

\bibitem[McClintock et al.(2003)]{mcc01B}McClintock, J. E., Narayan, R., Garcia, M. R., Orosz, J. A., Remillard, R. A., \& Murray S. S. 2003, ApJ, 593, 435 

\bibitem[McClintock \& Remillard(2003)]{mcc03}McClintock, J. E., \&
Remillard, R. A. 2003, astro-ph/0306213


\bibitem[Mirabel et al.(2001)]{Mir01}Mirabel, I. F., Dwawan, V., Mignani, R. P., Rodrigues, I., \& Guglielmetti, F. 2001, Nature, 413, 139

\bibitem[O'Donoghue \& Charles(1996)]{odo96}O'Donoghue, D., \& Charles, P. A. 1996, \mnras, 282, 191 


\bibitem[Orosz et al.(1995)]{oro95}Orosz, J. A. \& Bailyn C. D. 1995, \apj, 446, L59 

\bibitem[Orosz(2001)]{oro01}Orosz, J. A. 2001, ATEL~67 

\bibitem[Remillard et al.(2000)]{rem00}Remillard, R., Morgan, E., Smith, D., \& Smith, E. 2000, \iaucirc~7389
  
\bibitem[Rhodes \& Ruffini(1974)]{rod74}Rhodes, C. E., \& Ruffini, R. 1974, Phys. Rev. Lett., 32, 324      

\bibitem[Shahbaz et al.(1996)]{sha96}Shahbaz, T., Smale, A. P., Naylor, T., Charles, P. A., van Paradijs, J., Hassall, B. J. M., \& Callanan, P. 1996, \mnras, 282, 1437 

\bibitem[Soderblom et al.(1993)]{sod93}Soderblom, D. R., Stauffer,
J. R., Hudon, J. D., \& Jones, B. F. 1993, ApJS, 85, 315

\bibitem[Steeghs \& Stehle(1999)]{ste99}Steeghs, D., \& Stehle, R. 1999, \mnras, 307, 99

\bibitem[Tanaka \& Shibazaki(1996)]{tan96}Tanaka, Y., \& Shibazaki, N. 1996, ARA\&A, 34, 607

\bibitem[Torres et al.(2002a)]{tor02a}Torres, M. A. P.,  et al. 2002a, \apj, 569, 423

\bibitem[Torres et al.(2002b)]{tor02b}Torres, M. A. P.,  Casares, J.,
Mart\'\i{}nez-Pais, I. G., \& Charles P. A. 2002b, \mnras, 334, 233

\bibitem[Uemura(2000a)]{uem00a}Uemura, M., et al. 2000a, \iaucirc~7418

\bibitem[Uemura(2000b)]{uem00b}Uemura, M., et al. 2000b, \pasj, 52, 15L

\bibitem[van Paradijs \& McClintock(1995)]{van95}van Paradijs, J., \& McClintock, J. E. 1995, in X-ray Binaries, ed. W. H. G. Lewin, J. van Paradijs \& E. P. J. van den Heuvel. (Cambridge: Cambridge Univ. Press), 107

\bibitem[van Paradijs(1998)]{van98}van Paradijs, J. 1998, in The many Faces of Neutron Stars, ed. R. Buccheri, J. van Paradijs \& M. A. Alpar (Dordrecht: Kluwer), 279

\bibitem[Wagner et al.(2001)]{wag01}Wagner, R. M., Foltz, C. B., Shahbaz, T., Casares, J., Charles, P. A., Starrfield, S. G., \& Hewett, P. 2001, \apj, 556, 42

\bibitem[Warner(1995)]{war95}Warner, B. 1995, Cataclysmic Variables (Cambridge: Cambridge Univ. Press) 

\bibitem[Whitehurst \& King(1991)]{whi91}Whitehurst, R., \& King, A. J. 1991, \mnras, 249, 25

\bibitem[Zurita et al.(2002)]{zur02}Zurita, C.,  et al. 2002, \mnras, 333, 791


\end{thebibliography}
\end{document}